# Ultrafast hot carrier relaxation in silicon monitored by phase-resolved transient absorption spectroscopy

Martin Wörle[1], Alexander W. Holleitner[1,2], Reinhard Kienberger[1] and Hristo Iglev[1]

[1]Physik-Department, Technische Universität München, James-Frank-Straße 1, 85748 Garching, Germany
e-mail address: hristo.iglev@ph.tum.de

[2]Munich Center for Quantum Science and Technology (MCQST), Schellingstr. 4, 80799 München, Germany

The relaxation dynamics of hot carriers in silicon (100) is studied via a novel holistic approach based on phase-resolved transient absorption spectroscopy with few-cycle optical pulses. After excitation by a sub-5 fs light pulse, strong electron-phonon coupling leads to an ultrafast momentum relaxation with time constant of 10 fs. The thermalization of the hot carriers occurs on a time constant of 150 fs, visible in the temporal evolution of the collision time as extracted from the Drude model. We find an increase of the collision time from 3 fs for the shortest timescales with a saturation at approximately 18 fs. Moreover, the optical effective mass of the hot carrier ensemble evolves on ultrafast timescales as well, with a bi-exponential decrease from $0.7 \cdot m_e$ to about $0.125 \cdot m_e$ and time constants of 4 fs and 58 fs. The presented information on the electron mass dynamics as well as the momentum-, energy-, and collision-scattering times with unprecedented time resolution is important for all hot carrier optoelectronic devices.

Ultrafast scattering of highly energetic (hot) carriers in semiconductors plays crucial role in nanoscale semiconductor junctions, [1] hot carrier solar cells [2–4] and next-generation thermo- and optoelectronics [5,6]. The rapid development of time-resolved spectroscopy enables a direct monitoring of the scattering dynamics. In particular, time- and angle-resolved photoemission spectroscopy gives experimental information on the electron momentum and energy relaxation [7,8], while femtosecond THz spectroscopy is sensitive to hot-electron transport properties and to free-carrier plasma effects [9–11]. Furthermore, attosecond extreme ultraviolet (XUV) transient spectroscopy enables the access to the ultrafast dynamic of interband electron relaxation processes [12–14]. Despite accumulating knowledge, a clear understanding of the ultrafast scattering processes of hot electrons even in the conduction band of silicon (Si) is still incomplete, although it is the most fundamental semiconductor with significant importance for the microelectronic industry.

Recent progress in theoretical calculations from first principle has made it possible to gain a deeper insight into the electron relaxation dynamics in Si. The calculations show that the momentum scattering time due to electron-phonon coupling is supposed to be shorter than 10 fs for excess energies above 1.5 eV, although the value is one order of magnitude shorter than experimentally reported ones [8]. The concept of quasi-equilibrated hot-electron ensemble indicates that the transient relaxation of hot electrons is governed by two relaxation times with different magnitude: the momentum relaxation time $\tau_M$ and the energy relaxation time $\tau_E$ [7,8]. A very recent study using time-resolved photoemission spectroscopy and ab initio calculations [8] shows that the measured population decay of hot electrons corresponds to $\tau_E$, while the momentum relaxation rates are shown to be too short to be measured. Generally, transient reflective measurements and femtosecond THz spectroscopy on Si proved to be a powerful tool for monitoring Drude-response and related parameters, like effective carrier masses and collision times [10,15–17] as well as carrier-phonon interactions [9,11]. However, some of these experiments overestimate the extracted time constants due to limited time-resolution.

In this letter we report a peculiar phase-resolved transient absorption scheme with few-cycle pump and probe pulses. The comprehensive ansatz, with simultaneous detection of pump-induced absorption and refractive index changes, reveals a full picture of the ultrafast free carrier relaxation in Si, including the momentum relaxation time, energy relaxation time, transient effective carrier mass and collision time. The gained results could be important for novel application in electronics like energy storage [18], ultrafast electro-optical modulation techniques [19,20] and signal processing [21].

The principle of the current experiment is schematically shown in Figure 1. A collinear beam geometry at the sample was chosen to achieve the best possible temporal resolution (Figure 1a). The optical excitation is achieved by sub-5 fs pulses at 750 nm, while the transient relaxation dynamics are monitored in the spectral range between 1250 and 1950 nm. To get access to the pump introduced refractive index change the probe pulse is split into two co-propagating pulses (probe and reference). The temporal sequence of the used reference, pump and probe pulses is indicated in Figure 1b (for more details on the experiment see Supporting Information.) The energy transmittance $T$ of the probing pulse through the excited sample is measured for parallel (∥) and perpendicular (⊥) polarizations relative to the polarization plane of the pump and compared with the probe transmittance $T_0$ for blocked excitation beam. In this way, the induced change of the optical density, $\Delta OD_{\parallel,\perp} = -\log(T_{\parallel,\perp}/T_0)$ is determined for various probe wavelengths $\lambda$ and delay times $t_d$. These transient absorption measurements are performed with a blocked reference pulse.

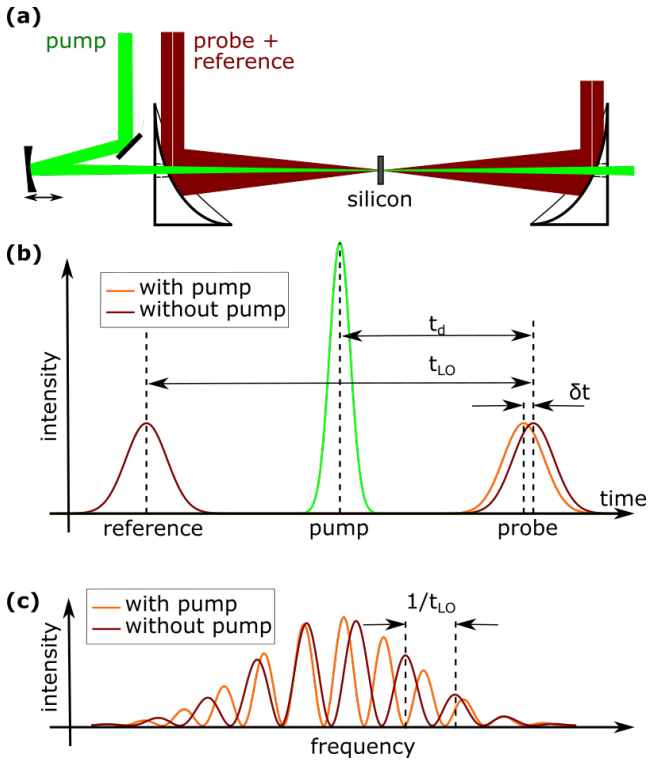

FIG. 1. (a) Schematic representation of the collinear sample layout. (b) Scheme of the used pulse sequence. The pump induced refractive index change causes a temporal shift of $\delta t$. (c) Schematics of the interference pattern measured by the IR detection system for both situations with and without excitation.

The overlap of probe and reference beams on the IR detector arrays leads to an interference pattern, which is schematically shown in Figure 1c. The frequency spacing between two neighboring maxima is inverse proportional to the reference-probe delay time ($\Delta v = 1/t_{LO}$), while the exact frequency position of the interference fringes depends on their relative phase shift. The photogenerated electrons and holes in the Si sample will change its refractive index $\Delta n$, indicated by the additional temporal shift $\delta t = \Delta n \cdot L/c$ in Figure 1b. Here $L$ is the sample thickness and $c$ denotes the speed of the light. Thus, $\Delta n$ causes a phase shift of the subsequent probe pulse relative to the reference pulse. Note that the reference passes through an unexcited sample, which implicates $t_d < t_{LO}$. The pump-induced phase shift $\Delta\Phi$ is extracted via Fourier-transform spectral interferometry [22,23]. In order to improve the signal-to-noise ratio, the spectrum measured with probe and reference beams is subtracted by the spectrum taken for blocked reference. The gained pure interference signal is converted from wavelength into frequency domain and Fourier transformed. The obtained peak at $1/t_{LO}$ is shifted to 0. The inverse Fourier transformation yields the phase difference between probe and reference. This procedure is repeated at every delay time $t_d$ for open and for blocked pump beam. The difference of the last two values provides the pump-induced phase shift $\Delta\Phi$. Note, that the phase-resolved measurements required sufficient modulation depth which additionally reduces the detectable spectral range to 1.35 – 1.85 µm.

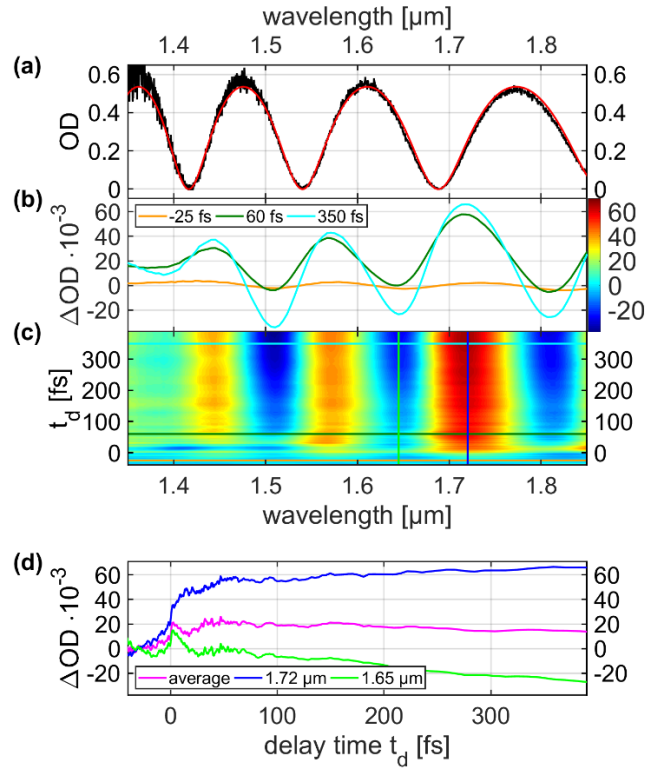

FIG. 2. (a) FTIR absorption spectra of a 2.5 µm thick Si (100) sample (experimental data in black, theory in red). (b) Isotropic transient absorption spectra measured at three selected delay times (see inset). (c) The full data set presented as a 2D color plot with color scale given next to b. (d) Signal transients measured at 1.65 µm and 1.72 µm, together with the signal averaged over the spectral range from 1.39 µm to 1.81 µm.

We emphasize, that the transient amplitude and phase changes yielded by this technique provide the full information of the pump-induced changes in the subsequent probing pulse [22,23]. Thus, the current method is comparable to the time-resolved THz spectroscopy [9–11]. Furthermore, the transfer from THz to optical probing frequencies facilitates a much better temporal resolution.

The investigated sample is a monocrystalline (100) silicon (Si) plate with a thickness $L \approx 2.5$ µm. The small sample thickness allows to overcome the limitations in the temporal resolution caused by the different group velocities of pump and probe pulses. However, the large refractive index of Si ($n_0 \approx 3.5$) leads to significant surface reflection of the probe radiation, which causes the thin sample plate to act as a Fabry-Pérot interferometer (FPI). The measured steady-state Fourier transform infrared (FTIR) absorption of the Si plate is shown in Figure 2a (black line). The cavity finesse extracted from the observed spectral modulations of 2.46 [24] is in excellent agreement with the expected surface reflection of 0.3 (see red fitted line in Fig. 2a). It should be noted, that averaging the steady-state spectrum over the free spectral range ($\Delta \nu_{FSR} = 2Ln_0/c$) of the FPI equals the sum of the individual attenuations of both cavity mirrors (i.e. $\bar{A} = -2\log_{10}(1-R)$).

The isotropic changes of the optical density ($\Delta OD_{iso} = (\Delta OD_{\parallel} + 2\Delta OD_{\perp})/3$) measured in the thin Si plate are shown in Figures 2b and 2c. The transient spectra measured for longer delay times (see cyan line in Figure 2b) show sinusoidal spectral modulations with positive and negative absorption values, which are off-phase compared to the steady-state spectrum shown in Figure 2 a. These observations give strong evidence for simultaneous pump-induced absorption and refractive index changes in the excited Si sample.

Averaging of the absorption data over a spectral range proportional to $\Delta \nu_{FSR}$ facilitates the extraction of the pure excited state population dynamics. The transient data averaged over the spectral range from 1.39 µm to 1.81 µm ($\approx 3 \times \Delta \nu_{FSR}$) are shown in Figure 2d (magenta), together with the transients measured at 1.65 µm and 1.72 µm. The figure indicates fast relaxation dynamics in the first 50 fs, followed by a slower signal increase in the next 500 fs. In general, the data measured in the absorption maximum at 1.72 µm shows similar temporal dynamics as the averaged signal, but at higher signal-to-noise ratio. This can be expected for samples with small spectral dynamics in the monitored spectral range.

The anisotropic absorption change $\Delta OD_{aniso} = (\Delta OD_{\parallel} - \Delta OD_{\perp})$ averaged over the spectral range from 1.39 µm to 1.81 µm are presented in Figure 3. The data indicate an extremely fast signal decay with a clear asymmetric temporal profile. The instrument response function of the setup (FWHM of $14 \pm 1$ fs) is shown for comparison. The anisotropic signal shows an additional exponential with decay time of $10 \pm 2$ fs (red line in Figure 3). The observation is a strong indication that the optical excitation with few-cycle pulses leads to an asymmetric population of the Si conduction band [11]. Taking the applied pumping energy of 1.8 µJ and the reflection losses on the front Si surface into account, the internal pumping intensity is estimated to be 0.17 TW/cm². These strong excitation fields lead to larger conduction band population in regions where the k-vector is parallel to the excitation field. The extracted time constant for the electron momentum relaxation of $\tau_M = 10 \pm 2$ fs is in agreement with the recently reported theoretical studies [8,25].

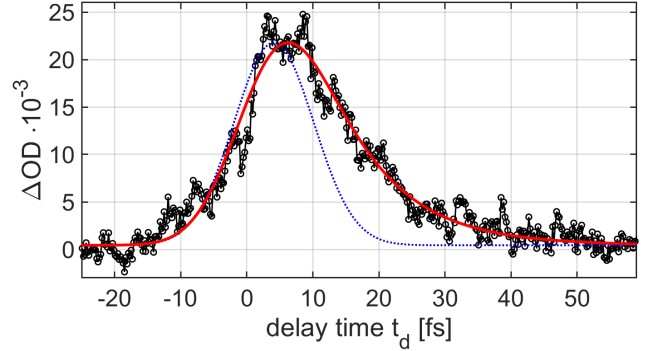

FIG. 3. Anisotropic (black) transient absorption changes averaged over the spectral range from 1.39 to 1.81 µm. The blue dotted illustrates the instrument response function.

Figure 4 presents the measured pump-induced isotropic phase shifts $\Delta \Phi_{iso} = (\Delta \Phi_{\parallel} + 2\Delta \Phi_{\perp})/3$, which reflects the transient change of the refractive index $\Delta n$. We emphasize, that the data shown in Figures 2 and 4 provide that the measured signals are in accordance to the expected negative $\Delta n$-values for free electrons. The time-resolved data show in-phase spectral modulation compared to the steady-state absorption (see Figure 2a), however, their relative modulation depth is much smaller than in the transient absorption signal (cf. Figure 2b). The transient phase shift shows an ultrafast initial signal change, which is followed by significant amplitude increase within the first 200 fs (see Figures 4b and c).

The simplest description of free carriers in response to an applied electric field in a photoexcited semiconductor is given by the Drude model. Optical-pump THz-probe spectroscopy on various system show the ability of this model to reproduce the measured data yielding information on the transient conductivity [26–28]. Assuming the validity of the Drude-model also in this high-field regime, the complex refractive index can be written as [29]:

$$\Delta \tilde{n} = -\frac{1}{2n_0}\left(\frac{e^2 N_{e-h}}{\varepsilon_0 m^*_{opt}}\right)\frac{1}{1 - 1/i\omega_{pr}\tau_c} \quad (1)$$

with the unperturbed refractive index $n_0$, electron charge $e$, density of photogenerated electron-hole pairs $N_{e-h}$, vacuum permittivity $\varepsilon_0$, optical effective mass of the charge carriers $m^*_{opt}$, central probe frequency $\omega_{pr}$ and the collision time $\tau_c$. By linking the real and imaginary parts of Equation 1 with the measured variables one gets [29]:

$$\Delta\Phi = -\frac{\pi L}{\lambda_{pr} n_0}\left(\frac{e^2 N_{e-h}}{\varepsilon_0 m^*_{opt}}\right)\frac{1}{\omega^2_{pr}+1/\tau^2_c}, \quad (2)$$

$$\Delta OD = \frac{\pi L}{\lambda_{pr} n_0 \ln 10}\left(\frac{e^2 N_{e-h}}{\varepsilon_0 m^*_{opt}}\right)\frac{2}{\omega_{pr}\tau_c(\omega^2_{pr}+1/\tau^2_c)}. \quad (3)$$

From Equations 2 and 3 the collision time can be extracted to

$$\tau_c = \frac{2|\Delta\Phi|}{\omega_{pr}\Delta OD \ln 10}. \quad (4)$$

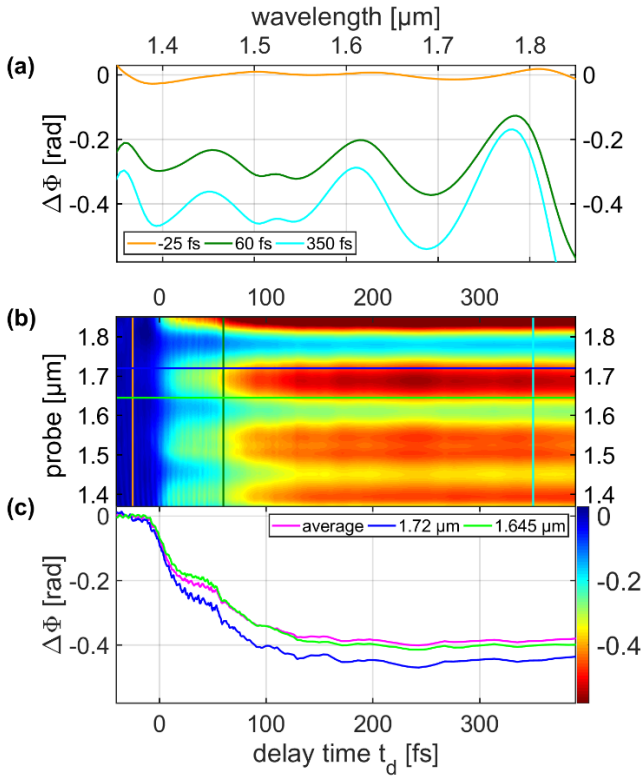

FIG. 4. Transient isotropic phase changes $\Delta\Phi_{iso}$ in Si after excitation with few-cycle laser pulse. (a) Phase changes measured at three selected delay times (see inset). (b) The full data set presented as a 2D color plot with color scale given next to c. (c) Data measured at 1.645 µm and 1.72 µm, together with the signal averaged over the spectral range from 1.39 µm to 1.81 µm.

The transient collision times calculated from the spectrally averaged phase shifts and absorption changes are presented in Figure 5a for probe polarization parallel (red) and perpendicular (gray) relative to the polarization plane of the pump. The significant difference between both data sets shows that the Drude model is only partially applicable in the first 30 fs. The reason for this phenomenon is the anisotropic population of the conduction band, which is reflected mainly in the transient absorption. Having said that, the data in Figure 5a indicate a very short collision time (< 3 fs) immediately after the optical excitation. This number agrees with theoretical studies on hot electrons in Si [30]. The short collision time comes mainly from elastic scattering of electrons from atoms [30]. The subsequent thermalization of the hot electron-hole gas reduces the carrier's kinetic energy. It can be expected that the decreased average velocity of the electrons in the conduction band will be reflected in an increase of $\tau_c$. In fact, Figure 5a shows a rise of the collision time from 3 fs to above 18 fs with a time constant of 150 ± 30 fs (see black line in Figure 5a). The extracted time constant is in accordance with the reported energy relaxation time of hot electrons of 100 – 120 fs [7,8]. It should be noted that according to the quadratic dependence of the excess energy on the carrier's velocity, the extracted relaxation time from Figure 5a could be a factor of 2 larger than the actual energy relaxation time.

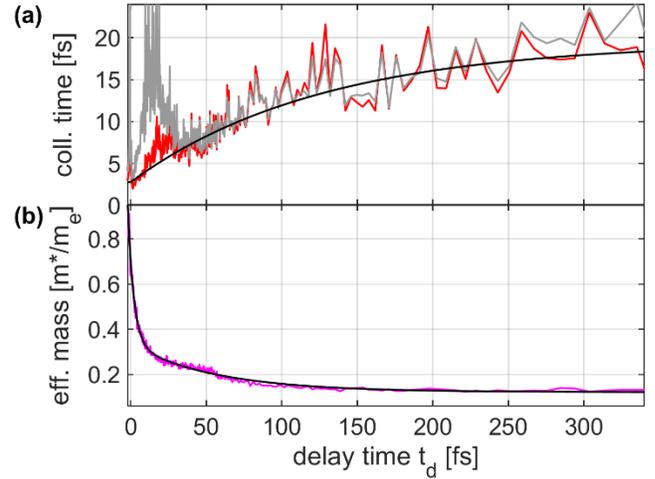

FIG. 5. (a) Collision time for probe polarization parallel (red) and perpendicular (gray) relative to the pump polarization. Calculated black line representing an exponential fit with a time constant of 150 ± 30 fs. (b) Relative effective mass of the charge carriers extracted from the spectral averaged experimental data (magenta). The black line represents a bi-exponential fit with time constants of 4 fs and 58 fs.

Assuming a constant carrier concentration on the sub-picosecond time scale and knowing the transient scattering time (shown in Figure 5a), Equation 2 enables to extract the effective relative carrier mass from the transient phase shift. The electron-hole pair concentration can be estimated from the averaged absorption data (see magenta line in Figure 2c, $\Delta OD \approx 0.016$ for $t_d = 250$ fs) and the reported electron-hole pair absorption cross-section $\sigma_{e-h} \approx 9.4 \cdot 10^{-18}$ cm$^2$ for a central wavelength of 1.6 μm [31]. The extracted electron-hole pair concentration is $N_{e-h} = (\ln 10 \cdot \Delta OD)/\alpha_{e-h} L \approx 1.6 \cdot 10^{19}$ cm$^{-3}$. The obtained transient evolution of $m^*_{opt}$ is presented in Figure 5b. For delay times $t_d > 250$ fs the effective mass reaches an almost constant value of about $0.125 \cdot m_e$. The latter value is in good agreement with the reported numbers for the 'optical' effective mass after thermalization of $0.156 - 0.168 \cdot m_e$ [17]. The deviation is most probably due to the uncertainty in the used effective carrier absorption cross section. However, we emphasize that our main focus was on the temporal evolution of the extracted effective carrier mass. The data in Figure 5b show a bi-exponential decay with faster decay time of 4 fs and slower time constant of $58 \pm 5$ fs.

The carrier's effective mass depends very much on the averaged energy distribution in the bands [30]. As discussed above, a strong pump pulse excites electrons at various $k$ points by tunnel and multiphoton excitations. Thus, the temporal evolution of the effective mass traces the averaged kinetic energy of the hot electron ensemble in the conduction band. The observed relaxation dynamics is in good agreement with the attosecond XUV transient absorption measurements performed by Schultze et al. [12], where a 5-fs time constant for purely electronic response and 60 fs for electron-lattice dynamics is reported. Note that coherent interactions between pump and probe pulses could affect the faster time constant.

In conclusion, we used a phase-resolved transient absorption spectroscopy with few-cycle pulses to study the relaxation dynamics of highly energetic (hot) carriers in silicon (100). The measured polarization resolved absorption data reveal an ultrafast momentum relaxation with a time constant of 10 fs. The simultaneous detection of refraction and absorption changes following the strong field sub-5 fs excitation provides a comprehensive information on the subsequent scattering dynamics. We applied the Drude model to extract the temporal evolutions of the carrier's effective mass and collision time. The latter increases from 3 fs to above 18 fs with a time constant of $150 \pm 30$ fs due to thermalization of the hot electron-hole gas. The extracted effective carrier mass indicates a bi-exponential decrease with time constants of 4 fs and 58 fs from almost 0.7 to about 0.125. We emphasize that detailed information on the hot electron transport properties is crucial for the applications in photovoltaic devices and next generation nano- and optoelectronics.

The authors acknowledge the financial support from the German Excellence Initiative via the Cluster of Excellence "e-Conversion" EXC 2089/1 – 390776260.